\documentclass[12pt]{iopart}
\usepackage{graphicx,epsfig}
\usepackage{amsthm}
\usepackage[latin2]{inputenc}

\newcommand{\ket}[1]{|#1\rangle}
\newcommand{\bra}[1]{\langle #1|}
\newcommand{\proj}[1]{\ket{#1}\bra{#1}}

\begin{document}

\title[]{Resource conversion between operational coherence and multipartite entanglement in many-body systems}

\author{Li-Hang Ren$^{1}$, Meng Gao$^{1}$, Jun Ren$^{1}$ and Z. D. Wang$^{2*}$ and Yan-Kui Bai$^{1, 2\dagger}$}
\address{$^{1}$College of Physics and Hebei Key Laboratory of Photophysics Research and Application, Hebei Normal University, Shijiazhuang, Hebei 050024, China}
\address{$^{2}$Department of Physics and HKU-UCAS Joint Institute for Theoretical and Computational Physics at Hong Kong, The University of Hong Kong, Pokfulam Road, Hong Kong, China}

\ead{zwang@hku.hk}
\ead{ykbai@semi.ac.cn}

\begin{abstract}
We establish a set of resource conversion relationships between quantum coherence and multipartite entanglement in many-body systems, where the operational measures of resource cost and distillation are focused. Under the multipartite incoherent operation, the initial coherence of  single-party system bounds the maximal amounts of corresponding operational entanglement in an arbitrary bipartite partition as well as the genuine multipartite entanglement in many-body systems. Furthermore, the converted multipartite entanglement can be transferred to its subsystems and restored to coherence of a single party by means of local incoherent operations and classical communication, which constitutes a protocol of resource interconversion within the full incoherent operation scenario. As an example, we present a scheme for cyclic interconversion between coherence and genuine multipartite entanglement in three-qubit systems without  loss. Moreover, we analyze the property of bipartite and multipartite genuine multi-level entanglement by the initial coherence and investigate multipartite resource dynamics in the conversion.
\end{abstract}


\maketitle

\section{Introduction}

Quantum entanglement \cite{epr35pr,bohr35pr}, as an important physical resource, has been widely applied to quantum communication and quantum computation, such as quantum teleportation, quantum key distribution, one-way quantum computation and so on (see a review paper \cite{hor08rmp}). In recent years, quantum coherence \cite{bcp14prl,abe06arxiv} has also been formulated as a physical resource in quantum information processing and becomes a vivid research topic \cite{sap17rmp,hhwpzf18pr}. Both kinds of the resources can be considered as the particular manifestation of the superposition principle in quantum mechanics, and their interplay has attracted a lot of attention. For example, the resource consumption in the tasks of quantum state preparation and quantum state merging \cite{how05nat} can be quantified by pairs of quantum coherence and entanglement \cite{chi16prl,str16prl}. Moreover, unified characterization and operational relations between the two kinds of resource measures were analyzed \cite{yao15pra,tan16pra,zhu17pra,zhu18pra,zhou19pra}.

It is a practical problem to investigate the resource conversion between quantum coherence and entanglement under certain operation constraints. Streltsov \textit{et al} first showed that single-party quantum coherence with respect to some fixed basis can be converted to entanglement via bipartite incoherent operations \cite{str15prl}. On the other hand, Chitambar \textit{et al} provided the upper bound of assisted coherence distillation (ACD) from bipartite systems to one of the subsystems by local quantum incoherent operations and classical communication (LQICC) \cite{csrbal16prl}. Other effective resource conversion methods for nonclassicality, quantum correlations and nonlocality were also presented \cite{ksp16prl,mygvg16prl,zhang19pra}, and some key experimental progress have been made in the optics and superconducting systems \cite{xiang17optica,xiang18prl,jin18pra,sun19prl}.

In the quantum resource theory (QRT) \cite{cg19rmp}, there are two basic operational processes: one is the so-called resource distillation, and the other is resource formation. Motivated by these processes, Winter and Yang established operational resource theory of coherence by focusing on the distillable coherence and coherence cost \cite{win16prl}, which correspond to the distillable entanglement \cite{ben96pra} and entanglement cost \cite{hht01jpa} in the QRT of entanglement. It is desirable and necessary to study the operational resource conversion between quantum coherence and entanglement, especially in multipartite systems. Moreover, the ACD in the LQICC scenario involves two parties (Alice and Bob) and their goal is to maximize the quantum coherence of Alice's subsystem by Bob performing arbitrary local quantum operations on his subystem, while Alice is restricted to local incoherent operations assisted by classical communication between them \cite{csrbal16prl}. It is noted that the LQICC can generate coherence in Bob's subsystem since local quantum operations are not incoherent operations in general \cite{str17prx}. In a cyclic resource conversion of coherence-entanglement-coherence, the operations should be considered within the full incoherent scenario, \textit{i.e.}, both Alice and Bob are restricted to local incoherent operations which is referred to as local incoherent operations and classical communication (LICC). The definitions can be further generalized to multipartite systems and we have the relation  LICC $\subset$ LQICC \cite{str17prx}.
In addition, since entanglement in many-body systems is a complex problem, it is meaningful to characterize the entanglement property via quantum coherence in the resource conversion. Meanwhile, the resource dynamics of quantum coherence and multipartite entanglement in the conversion needs to be addressed, because quantum systems interact unavoidably with the environment in realistic quantum information processing.

\begin{figure}
	\begin{center}
		\epsfig{figure=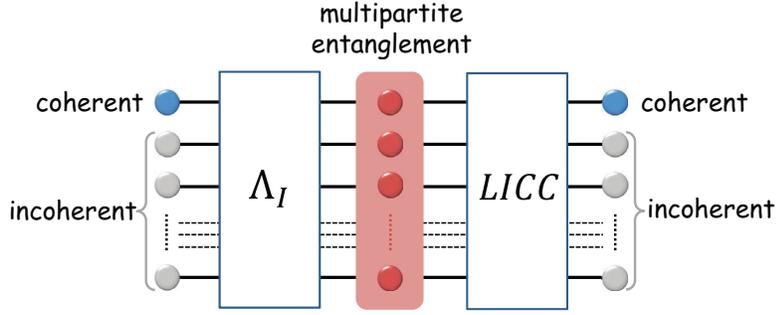,width=0.65\textwidth}
	\end{center}
	\caption{(Color online) A conceptual diagram for the cyclic operational resource conversion between single-party quantum coherence and multipartite entanglement within the full incoherent operation scenario, where $\Lambda_I$ is a global multipartite incoherent operation and the LICC means local incoherennt operations and classical communication.}\label{f1}
\end{figure}

In this paper, focused on resource distillation and resource formation, we study the conversion between quantum coherence and multipartite entanglement in many-body systems, which is different from the previous results in bipartite quantum systems \cite{tan16pra,zhu17pra,zhu18pra,zhou19pra,str15prl}.
As shown in Fig. \ref{f1}, we give a conceptual diagram for the cyclic resource conversion within the full incoherent operation scenario.
In the conversion from coherence to entanglement, we restrict the operation to multipartite incoherent operations which are chosen to be a kind of global operations \cite{yao15pra}. Moreover, in the conversion from multipartite entanglement to single-party coherence, we utilize the local incoherent operations and classical communication (LICC) \cite{str17prx} which does not generate coherence in the assisted subsystems and can overcome the flaw of local quantum incoherent operations and classical communication (LQICC) in the cyclic resource conversion.
It is found that, via multipartite incoherent operations, the initial operational coherence of single party bounds not only the generated bipartite operational entanglement but also the genuine multipartite entanglement in the composite systems. The converted multipartite entanglement can be further transferred to its subsystems and restored to the coherence of a single-party subsystem by the LICC, where we prove that the optimal resource conversion can make the relations saturated. As a typical application, we present a cyclic interconversion protocol between quantum coherence and multipartite entanglement in three-qubit systems without loss. In addition, we show that the coefficients of initial single-party coherent state can determine whether the converted quantum state is genuine multi-level entangled. Finally, under the depolarization  environment, different resource dynamical properties are investigated in multipartite resource conversion.

\section{Operational resource conversion in multipartite systems}

In the quantum resource theory, a free state is the one without possessing a defined resource, and free operation cannot generate the resource and maps the set of free states onto itself. In regard to the operational resource characterization, resource distillation is a transformation from a mixed state to the unit resource, and resource formation is the reverse transformation from the unit resource to a mixed state, where both of the transformations are restricted to free operations. For entanglement theory, two motivated measures arising from the two operational tasks are distillable entanglement and entanglement cost \cite{hor08rmp},
\begin{eqnarray}
E_d(\rho)&=&\mbox{sup}\{r:\lim\limits_{n \to \infty}[\inf\limits_{\Gamma}||\Gamma(\rho^{\otimes n})-\Phi^{\otimes rn}_{2+}||_1]=0\},\\
E_c(\rho)&=&\mbox{inf}\{r:\lim\limits_{n \to \infty}[\inf\limits_{\Gamma}||\rho^{\otimes n}-\Gamma(\Phi^{\otimes rn}_{2+})||_1]=0\},
\end{eqnarray}
where $r$ is the optimal rate in the tasks, $||\cdot||_1$ trace norm, $\Phi_{2+}=\proj{\Phi_{2}^{+}}$ the two-qubit Bell state $\ket{\Phi_2^+}=(\ket{00}+\ket{11})/\sqrt{2}$ (unit entanglement), and $\Gamma(\cdot)$ the free operation in resource theory of entanglement [\textit{i.e.}, local operations and classical communication (LOCC) \cite{loccnote}]. It is noted that entanglement cost is not equal to the well-known entanglement of formation $E_f(\rho)$ in general \cite{has09nphy,fuk07jmp}.

In the coherence resource theory formulated by Baumgratz \textit{et al} \cite{bcp14prl}, the free state has the form $\sigma=\sum_ip_i\proj{i}$ with $\{\ket{i}\}$ being a fixed reference basis in finite dimensions, and the set of incoherent states is denoted by $\mathcal{I}$. The incoherent operation is the free operation, which is specified by a set of Kraus operators $\{K_l\}$ satisfying $\sum_lK_l^\dagger K_l=I$ and $ K_l\mathcal{I} K_l^\dagger\subset \mathcal{I}$ for all $l$.
The unit coherence resource is the maximal coherent single-qubit state $\Psi_2=\frac{1}{2}\sum_{i,j=0}^{1}\ket{i}\bra{j}$. The framework can be further generalized to multipartite scenario, in which the $N$-partite incoherent state has the form \cite{yao15pra}
$\sigma=\sum_{\vec{i}} p_{\vec{i}} \proj{\vec{i}}$ where $\ket{\vec{i}}=\ket{i_1}\otimes \ket{i_2}\otimes \cdots \ket{i_n}$ with $\ket{i_k}$ being a pre-fixed local basis of the $k$th subsystem. The $N$-partite incoherent operation can also be expressed by a set of Kraus operators, where each incoherent operator maps the set of $N$-partite incoherent state onto itself (see the details in Appendix A).

The coherence distillation is the process that extracts unit coherence from a mixed state by incoherent operations, and the dual coherence formation is a process that prepares a mixed state by consuming unit coherent states under incoherent operations. In the asymptotic limit of many copies of a state, it is shown that distillable coherence and coherence cost can be quantified by simple single-letter formulas \cite{win16prl}
\begin{eqnarray}
C_d(\rho)&=&C_r(\rho)=S(\Delta(\rho))-S(\rho),\\
C_c(\rho)&=&C_f(\rho)=\mbox{min}\sum_ip_i S(\Delta(\psi_i)),
\end{eqnarray}
where $C_r(\rho)$ is the relative entropy of coherence \cite{bcp14prl} with $S(\rho)=-\tr\rho\mbox{log}_2\rho$ being the von Neumann entropy and $\Delta(\rho)=\sum_i \bra{i}\rho\ket{i}\proj{i}$, and $C_f(\rho)$ is the coherence of formation with the minimum running over all the pure state decompositions $\rho=\sum_i p_i\proj{\psi_i}$.

\subsection{Converting operational coherence to multipartite entanglement via multipartite incoherent operations}

Quantum coherence and entanglement are two kinds of fundamental nonclassical resources that can each be characterized within an operational resource theory. It is desirable to study the operational resource conversion between them. The conversion from quantum coherence to entanglement via bipartite incoherent operations was first studied by Streltsov \textit{et al},
where two resources are quantified by distance-based measures such as relative entropy and Uhlmann fidelity \cite{str15prl}. Based on bipartite incoherent operations, Zhu \textit{et al} further gave the conversion relationship between coherence and entanglement quantified by measures of convex-roof extension, $l_1$-norm of coherence and negativity \cite{zhu17pra,zhu18pra}.

It should be noted that the multipartite incoherent operation is not equivalent to that of the bipartite case in general \cite{yao15pra,str15prl}. For example, the bipartite incoherent operation can convert a single maximal coherent qubit  plus an incoherent ancila to the Bell state, but the tripartite incoherent operation makes the coherent qubit  plus two incoherent qubits into the GHZ state, which means that the multipartite operation has the ability to generate multipartite entangled states (see the details in the last paragraph of Appendix A). Here, motivated by resource distillation and recourse cost, we study the operational resource conversion between coherence and entanglement in multipartite systems, and can obtain the following theorem.

\textit{Theorem 1}. Given a multipartite incoherent operation $\Lambda_{I}$ applied to a single-party coherent state $\rho_{A}$ and an ancillary $N$-partite incoherent state $\sigma_{B_1B_2\dots B_n}$, the generated operational entanglements are upper bounded by the operational coherences of single-party system
\begin{eqnarray}
C_d(\rho_A)&\geq& E_d\left[\Lambda_I(\rho_{A}\otimes \sigma_{B_1B_2\dots B_n})\right], \\
C_c(\rho_A)&=&C_f(\rho_A) \geq \mbox{max}\{E_c[\Lambda_I(\rho_{A}\otimes \sigma_{B_1B_2\dots B_n})], \nonumber\\
&& E_f[\Lambda_I(\rho_{A}\otimes \sigma_{B_1B_2\dots B_n})]\},
\end{eqnarray}
where the coherent state has the form $\rho_{A}=\sum_{mn}\rho_{mn}\ket{m}\bra{n}$ and the operational entanglements are in an arbitrary bipartite partition $\alpha|\bar{\alpha}$ for the multipartite systems with $\alpha\bigcup \bar{\alpha}=\{AB_1B_2\dots B_n\}$.

\textit{Proof}.---
We first prove the relation in Eq. (5). Since the distillable coherence is equal to the relative entropy of coherence \cite{win16prl}, we have
\begin{eqnarray}
	C_d(\rho_{A})&=&S(\rho_{A}\|\sigma_{A}) \nonumber\\
	& =&S(\rho_{A}\otimes\sigma_{B_1B_2\dots B_n}\|\sigma_{A}\otimes\sigma_{B_1B_2\dots B_n}) \nonumber\\
	&  \geq& S[\Lambda_{I}(\rho_{A}\otimes\sigma_{B_1B_2\dots B_n})\|\Lambda_{I}(\sigma_{A}\otimes\sigma_{B_1B_2\dots B_n})] \nonumber\\
	& \geq& E_r(\Lambda_{I}(\rho_{A}\otimes\sigma_{B_1B_2\dots B_n})) \nonumber\\
	& \geq& E_d(\Lambda_{I}(\rho_{A}\otimes\sigma_{B_1B_2\dots B_n})),
\end{eqnarray}
where $\sigma_A$ is the closest incoherent state to $\rho_A$ in the first equation, the additive and contractive properties of relative entropy \cite{ved97prl,ved02rmp} are used in the second equation and the first inequality, the result of the second inequality comes from the definition of the relative entropy of entanglement, and the last inequality is due to the fact that the relative entropy of entanglement is the upper bound of distillable entanglement \cite{hhh00prl,dw05prsa}. In the first inequality of the proof, the quantum state $\Lambda_{I}(\sigma_{A}\otimes\sigma_{B_1B_2\dots B_n})$ is multipartite incoherent which is also $(N+1)$-partite separable. Therefore, the distance of relative entropy in the first inequality is not less than the bipartite relative entropy of entanglement $E_r$ in the second inequality where the separable state can be chosen in an arbitrary bipartite partition $\alpha|\bar{\alpha}$ such as $A|B_1B_2\cdots B_n$, $AB_1|B_2\cdots B_n$, $AB_1B_2|\cdots B_n$ and so on, which further results in the last inequality in Eq. (7) being satisfied for the corresponding bipartite partition.
Then, we prove the second relation  shown in Eq. (6). Because the equality $C_c(\rho_A)=C_f(\rho_A)$ and the additivity of coherence of formation, we can obtain
\begin{eqnarray}
	C_c(\rho_{A})&=&C_f(\rho_{A}\otimes\sigma_{B_1B_2\dots B_n}) \nonumber\\
	& \geq& C_f\left[\Lambda_I(\rho_{A}\otimes\sigma_{B_1B_2\dots B_n})\right] \nonumber\\
	&  =& \lim\limits_{n\rightarrow\infty} \frac{1}{n}C_f\left[\Lambda_I(\rho_{A}\otimes\sigma_{B_1B_2\dots B_n})^{\otimes n}\right] \nonumber\\
	& \geq&\lim\limits_{n\rightarrow\infty} \frac{1}{n}E_f[\Lambda_I(\rho_{A}\otimes\sigma_{B_1B_2\dots B_n})^{\otimes n}] \nonumber\\
	& =& E_c[\Lambda_I(\rho_{A}\otimes \sigma_{B_1B_2\dots B_n})],
\end{eqnarray}
where the monotone property of $C_f$ under incoherent operations is used in the first inequality, the second inequality holds due to the property $C_f(\varrho)\geq E_f(\varrho)$ \cite{zhu17pra} with the entanglement quantified in an arbitrary bipartite partition $\alpha|\bar{\alpha}$ for composite systems $AB_1B_2\cdots B_n$, and the last equation comes from the relation that entanglement cost is equal to the regularized entanglement of formation \cite{hht01jpa} with the bipartite partition for $E_c$ being arbitrary. Moreover, for the case of a single copy of coherent state $\rho_A$, we can derive that $C_f(\rho_A)\geq E_f[\Lambda_I(\rho_{A}\otimes \sigma_{B_1B_2\dots B_n})]$ by a similar analysis and the bipartite partition for $E_f$ is also arbitrary.
Although the two operational entanglements $E_c(\varrho)\neq E_f(\varrho)$ in general \cite{fuk07jmp}, we can get that, given a multipartite incoherent operation $\Lambda_I$, the coherence of formation $C_f(\rho_A)$ is not less than the maximum of operational entanglements for any bipartite partition in the multipartite system $AB_1B_2\cdots B_n$ and then Eq. (6) is satisfied, such that we complete the proof of the theorem. \qed

Compared with bipartite entanglement, the characterization of multipartite entanglement is much more complicated. In the pure state case, a multipartite quantum state is genuine multipartite entangled if it cannot be written as a bipartite product state under any bipartite partitions \cite{dur99prl,vice11pra,huber14prl}. For the quantification of entanglement in many-body systems, there exist a kind of genuine multipartite entanglement (GME) measures generalized by bipartite entanglement measures \cite{dai20prapp,sza15pra},
\begin{equation}
E^{GME}(\ket{\psi})=\mbox{min}_{\{\alpha\}} E_\alpha(\ket{\psi}),
\end{equation}
where $\ket{\psi}$ is an $N$-partite pure state and $\alpha$ represents all possible bipartite partitions $\alpha|\bar{\alpha}$ in the composite systems. When the bipartite measure $E$ is chosen to be the operational entanglements $E_d$ and $E_f$, we can obtain the GME measures $E_d^{GME}$ and $E_f^{GME}$. In Theorem 1, when the initial coherent state and final output state after the operation $\Lambda_I$ are pure states, we can get the following conversion relations.

\textit{Corollary 1}. In the operational resource conversion between single-party coherence and multipartite entanglement, the GME measures quantified via $E_d$ and $E_f$ are upper bounded by the operational coherences,
\begin{eqnarray}
C_d(\ket{\phi}_A)&\geq& E_d^{GME}(\ket{\psi}_{AB_1B_2\cdots B_n}),\\
C_f(\ket{\phi}_A)&\geq& E_f^{GME}(\ket{\psi}_{AB_1B_2\cdots B_n}),
\end{eqnarray}
where $\ket{\phi}_A$ is the initial coherent pure state and $\ket{\psi}_{AB_1B_2\cdots B_n}$ is the output state under the multipartite incoherent operation $\Lambda_I(\proj{\phi}_A\otimes \sigma_{B_1B_2\cdots B_n})$ with an $N$-partite incoherent state being ancillary.

\textit{Proof.}--- According to Theorem 1, we know that the single-party distillable coherence $C_d(\ket{\phi}_A)$ is the upper bound on the distillable entanglement $E_d(\ket{\psi}_{AB_1B_2\cdots B_n})$ in an arbitrary bipartite partition of the multipartite systems. Therefore, $C_d(\ket{\phi}_A)$ is not less than $E_{d}(\ket{\psi}_{\alpha|\bar{\alpha}})$ which is the minimal bipartite distillable entanglement, and then we can obtain the inequality in Eq. (10) based on the definition of $E_d^{GME}$. The situation for coherence of formation $C_f(\ket{\phi}_A)$ is similar and we can derive the inequality in Eq. (11) after analyzing the relation between $E_f^{GME}$ and $E_f$ in multipartite systems.  \qed

The GME measure in Eq. (9) can be further generalized to mixed states by the convex roof extension \cite{dai20prapp}. A multipartite mixed state is said to be genuine multipartite entangled if, in any pure state decomposition of the mixed state, there exists at least one pure state component which cannot be written as bipartite product state with respect to any bipartite partition.  For example, the GME  based on entanglement of formation $E_f$ can be quantified by
\begin{equation}
E_f^{GME}(\rho)=\mbox{inf}_{\{p_i,\ket{\psi_i}\}} \sum_i p_i E_f^{GME}(\ket{\psi_i}),
\end{equation}
where the minimum runs over all the pure state decompositions $\rho=\sum_i p_i \ket{\psi_i}\bra{\psi_i}$. Although the GME measure for mixed states can be constructed by distillable entanglement, we do not utilize $E_d^{GME}(\rho)$ to characterize the multipartite entanglement in the operational resource conversions since $E_d$ is not a measure based on the convex roof construction of pure state measure. Here we consider the multipartite relative entropy of entanglement which has the form \cite{modi10prl}
\begin{equation}
E_r^M(\rho)=\min\limits_{\sigma_s\in \mathcal{D}} S(\rho||\sigma_s),
\end{equation}
where $S(x||y)\equiv-\mbox{tr}(x\mbox{log}y)+\mbox{tr}(x\mbox{log}x)$ is the relative entropy with $\sigma_s=\sum_i p_i \proj{\varphi_1^i}\otimes \proj{\varphi_2^i}\otimes\cdots\otimes\proj{\varphi_n^i}$ being the element of the $N$-partite separable state set $\mathcal{D}$. In the resource conversion of mixed states, we have the following theorem.

\textit{Theorem 2}. Given a multipartite incoherent operation $\Lambda_I$ applied to a single-party coherent state $\rho_A$ and an ancillary $N$-partite incoherent state $\sigma_{B_1B_2\cdots B_n}$, the generated multipartite entanglements are upper bounded by the operational coherences of initial quantum state
\begin{eqnarray}
C_d(\rho_A)&\geq& E_r^M[\Lambda_I(\rho_A\otimes\sigma_{B_1B_2\cdots B_n})],\\
C_f(\rho_A)&\geq& E_f^{GME}[\Lambda_I(\rho_A\otimes\sigma_{B_1B_2\cdots B_n})],
\end{eqnarray}
where $E_r$ is $(N+1)$-partite relative entropy of entanglement and $E_f^{GME}$ is the GME measure based on entanglement of formation.

\textit{Proof.}---For the single-party quantum state $\rho_A$, its distillable coherence $C_d$ is equal to the relative entropy of coherence $C_r$, and we have
\begin{eqnarray}
C_d(\rho_A)&=&C_r(\rho_A)=S(\rho_A||\sigma_A)\nonumber\\
&\geq&S[\Lambda_I(\rho_A\otimes\sigma_{B_1B_2\cdots B_n})||\Lambda_I(\sigma_A\otimes\sigma_{B_1B_2\cdots B_n})]\nonumber\\
&\geq&E_r^M[\Lambda_I(\rho_A\otimes\sigma_{B_1B_2\cdots B_n})],
\end{eqnarray}
where the additive and contractive properties of relative entropy are used in the first inequality, and the second inequality is satisfied due to $\Lambda_I(\sigma_A\otimes\sigma_{B_1B_2\cdots B_n})$ being an $(N+1)$-partite incoherent state which is a multipartite separable state. For the coherence of formation, we can obtain
\begin{eqnarray}
C_f(\rho_A)&\geq& C_f[\Lambda_I(\rho_{A}\otimes \sigma_{B_1\dots B_n})]\nonumber\\
&=&\sum_i p_i C_f(\ket{\psi_i^\prime}_{AB_1B_2\cdots B_n})\nonumber\\
&\geq&\sum_i p_i E_f(\ket{\psi_i^\prime}_{\alpha|\bar{\alpha}})\nonumber\\
&\geq&\sum_i p_i E_f^{GME}(\ket{\psi_i^\prime}_{AB_1B_2\cdots B_n})\nonumber\\
&\geq&E_f^{GME}[\Lambda_I(\rho_{A}\otimes \sigma_{B_1\dots B_n})],
\end{eqnarray}
where the additive and contractive properties are used in the first inequality, the optimal pure state decomposition $\Lambda_I(\rho_{A}\otimes \sigma_{B_1\dots B_n})=\sum_i p_i \proj{\psi_i^\prime}_{AB_1B_2\cdots B_n}$ for coherence of formation is utilized in the next equality, the second inequality holds due to the property $C_f(\ket{\psi_i^\prime})\geq E_f(\ket{\psi_i^\prime})$ with the entanglement in an arbitrary bipartite partition $\alpha|\bar{\alpha}$, the last two inequalities come from the definitions of multipartite entanglement $E_f^{GME}$ for pure states and mixed states, and then we complete the proof. \qed

Entanglement monogamy is an important property in many-body quantum systems, which gives the trade-off relations on the distribution of entanglement among different subsystems \cite{ckw00pra,koa04pra,osb06prl,kim10jpa,bxw14prl}. As is known, the monogamy property can be used for constructing multipartite entanglement measures and indicators \cite{ckw00pra,bxw14prl,oaf07pra,bai07pra,bai09pra,bxw14pra}. For the operational entanglement measures in an $N$-qubit quantum state $\rho_N=\rho_{A_1A_2\cdots A_n}$, we can define two multipartite entanglement indicators
\begin{eqnarray}
\tau_{MED}(\rho_{N})&=&\{\max[0,E_d^2(\rho_{A_1|\bar{A_1}})-\sum_{i=2}^n E_d^2(\rho_{A_1A_i})]\}^{\frac{1}{2}},\\
\tau_{MEF}(\rho_{N})&=&[E_f^2(\rho_{A_1|\bar{A_1}})-\sum_{i=2}^n E_f^2(\rho_{A_1A_i})]^{\frac{1}{2}},
\end{eqnarray}
where $\bar{A_1}=A_2A_3\cdots A_n$ is the subsystem other than qubit $A_1$. The indicators $\tau_{MED}$ and $\tau_{MEF}$ can indicate multipartite entanglement which cannot be stored in two-qubit subsystems. The residual entanglement in $\tau_{MEF}$ is always nonnegative according to the monogamy property of squared entanglement of formation \cite{bxw14prl,bxw14pra}. In Apendix B, we show that the distillable entanglement is monogamous in N-qubit pure states and some kinds of mixed states, which can be used to indicate the existence of the genuine multi-qubit entanglement when $\tau_{MED}$ is positive. In the operational resource conversion between coherence and multi-qubit entanglement, we have the following theorem.

\textit{Theorem 3}. For a single-qubit coherent state $\rho_A$ accompanied by an $N$-qubit incoherent state $\sigma_{B_1B_2\cdots B_n}$, the generated multi-qubit entanglements via multipartite incoherent operation $\Lambda_I$ are upper bounded by operational coherences of initial quantum state
\begin{eqnarray}
C_d(\rho_A)&\geq& \tau_{MED}[\Lambda_I(\rho_A\otimes \sigma_{B_1B_2\cdots B_n})], \\
C_f(\rho_A)&\geq& \tau_{MEF}[\Lambda_I(\rho_A\otimes \sigma_{B_1B_2\cdots B_n})],
\end{eqnarray}
where  $\tau_{MED}$ and $\tau_{MEF}$ are multi-qubit entanglement indicators based on distillable entanglement and entanglement of formation.

\textit{Proof.}--- We first consider the distillable coherence and multi-qubit entanglement indicator $\tau_{MED}$ and set $\rho_{AB_1B_2\cdots B_n}=\Lambda_I(\rho_{A}\otimes \sigma_{B_1B_2\dots B_n})$. According to Theorem 1, we have
\begin{eqnarray}
C_d(\rho_A)&\geq& E_d(\rho_{A|B_1B_2\cdots B_n})\nonumber\\
&\geq& [\max\{0,E_d^2(\rho_{A|B_1B_2\cdots B_n})-\sum_{i=1}^n E_d^2(\rho_{AB_i})\}]^{\frac{1}{2}}\nonumber\\
&=& \tau_{MED}(\rho_{AB_1B_2\cdots B_n}),
\end{eqnarray}
where in the first inequality we choose the bipartite partition $A|B_1B_2\cdots B_n$ for the distillable entanglement,
and in the second inequality we use the nonnegative property of two-qubit distillable entanglements. For the coherence of formation,
we can obtain
\begin{eqnarray}
C_f(\rho_A) &\geq& E_f(\rho_{A|B_1B_2\cdots B_n})\nonumber\\
&\geq& [E_f^2(\rho_{A|B_1B_2\cdots B_n})-\sum_{i=1}^n E_f^2(\rho_{AB_i})]^{\frac{1}{2}}\nonumber\\
&=& \tau_{MEF}(\rho_{AB_1B_2\cdots B_n}),
\end{eqnarray}
where the first inequality holds due to Eq. (6) in Theorem 1, and the second inequality comes from the monogamy property of squared entanglement of formation, which completes the proof of this theorem.  \qed

Now we consider the optimal conversion from the coherence to quantum entanglement via multipartite incoherent operations. It was shown that the generalized controlled-not operation is the optimal bipartite incoherent operation in bipartite entanglement  conversion \cite{str15prl}, when the ancilla is in the zero state $\ket{0}$. The multipartite generalized controlled-not operation $U_{mcn}$ can realize the transformation
\begin{equation}
\ket{i}\ket{j_1}\dots\ket{j_n}\rightarrow \ket{i}\ket{(i+j_1)\mbox{mod}~d}\dots \ket{(i+j_n)\mbox{mod}~d},
\end{equation}
which is the optimal multipartite incoherent operation $\Lambda_I$ in the operational resource conversion described by the above theorems when the state of the ancilla is $\sigma_{B_1\dots B_n}=\proj{0}^{\otimes n}$.
In Appendix A, we first show that the operation $U_{mcn}$ is a multipartite incoherent operation different from that of the bipartite case, and then prove its optimality for the multipartite resource conversion in Theorem 1 and its corollary. Under the optimal multipartite incoherent operation, the generated multipartite quantum state is $U_{mcn}[\rho_A\otimes\proj{0}^{\otimes n}_{B_1\cdots B_n}]U_{mcn}^\dagger$ and can be written as
\begin{equation}
\rho_{AB_1B_2\dots B_n}^{mc}=\sum_{m,n=0}^{d-1} \rho_{mn}\ket{mm\ldots m}\bra{nn\ldots n},
\end{equation}
which is the so-called maximally correlated state (MCS) \cite{rai99pra} and is multipartite entangled with all its reduced states being separable.
For the case of MCS, we have the desired properties $E_d(\rho^{mc})=E_r(\rho^{mc})$ \cite{rai01ieee} and $E_c(\rho^{mc})=E_f(\rho^{mc})$ \cite{vdc02prl} in an arbitrary bipartite partition of multipartite systems   $AB_1B_2\dots B_n$, which saturate the inequalities in Theorem 1 and make the resource conversions optimal. Since the saturated equalities in Theorem 1 are satisfied in any bipartite partition, we can get that the conversions between coherence and multipartite entanglement in Corollary 1 are also optimal according to the definition of $E^{GME}$.  Moreover, the multipartite generalized controlled-not operation $U_{mcn}$ is also optimal for multipartite resource conversions described by Theorem 2 and Theorem 3, which makes the corresponding inequalities saturated and we give the detailed proofs in Appendix C.

\subsection{Resource conversion from multipartite entanglement to coherence via LICC}
It is useful to establish the operational relations between entanglement and quantum coherence. Zhu \textit{et al} proved a one-to-one mapping between the two kinds of resource measures based on convex roof extension, and showed the relation $E_f(\rho_{AB})\leq C_f(\rho_{AB})$ in bipartite quantum states \cite{zhu17pra}. For multipartite systems, we can obtain the following theorem for operational resource measures.

\textit{Theorem 4}. In an $N$-partite quantum state $\rho_{N}=\rho_{A_1A_2\cdots A_n}$, the operational entanglements and quantum coherence are connected by the relations
\begin{eqnarray}\label{26}
E_d(\rho_{\alpha|\bar{\alpha}})&\leq&E_r^M(\rho_{N})\leq C_d(\rho_{N}),\\
E_f^{GME}(\rho_N)&\leq& E_f(\rho_{\alpha|\bar{\alpha}})\leq C_f(\rho_{N}),
\end{eqnarray}
where $\alpha|\bar{\alpha}$ is an arbitrary bipartite partition in the composite system, and the  bounds are saturated by the maximally correlated states.

\textit{Proof.}--- For the $N$-partite quantum state, its distillable entanglement in the partition $\alpha|\bar{\alpha}$
is not greater than the relative entropy of entanglement $E_d(\rho_{\alpha|\bar{\alpha}})\leq E_r(\rho_{\alpha|\bar{\alpha}})$ \cite{hhh00prl,dw05prsa}. According to the definitions of bipartite and multipartite relative entropy of entanglements, we have $E_r^M(\rho_N)=\mbox{min}_{\{\sigma_s\in \mathcal{D}\}}S(\rho_N||\sigma_s)=S(\rho_N||\sigma_s^N)= S(\rho_{\alpha|\bar{\alpha}}||\sigma_s^{\alpha|\bar{\alpha}})\geq E_r(\rho_{\alpha|\bar{\alpha}})$, in which $\sigma_s^N=\sigma_s^{A_1A_2\cdots A_n}$ is the nearest $N$-partite separable state for $E_r^M$ and $\sigma_s^{\alpha|\bar{\alpha}}$ is the bipartite partition expression of $\sigma_s^N$. Since the set of $N$-partite incoherent states is a subset of $N$-partite separable states, we can get $E_r^M(\rho_N)\leq C_r(\rho_N)=C_d(\rho_N)$, and then the inequality in Eq.(26) is satisfied. For the optimal pure state decomposition of $\rho_{\alpha|\bar{\alpha}}$, the bipartite entanglement of formation has the property
$E_f(\rho_{\alpha|\bar{\alpha}})=\sum_i p_i E_f(\ket{\psi_{\alpha|\bar{\alpha}}^i})\geq \sum_i p_i E_f^{GME}(\ket{\psi_{N}^i})\geq E_f^{GME}(\rho_N)$. Combining this property with the relation $C_f(\rho_N)\geq E_f(\rho_{\alpha|\bar{\alpha}})$, we have the inequality in Eq.(27). Moreover, when the $N$-partite quantum state is the MCS $\rho_N^{mc}=\sum p_{mn}\ket{mm\cdots m}\bra{nn\cdots n}$, we have $E_d(\rho_{\alpha|\bar{\alpha}}^{mc})=E_r^M(\rho_{N}^{mc})= C_d(\rho_{N}^{mc})$ and $E_f^{GME}(\rho_N^{mc})= E_f(\rho_{\alpha|\bar{\alpha}}^{mc})=C_f(\rho_{N}^{mc})$, which completes the proof. \qed

Next, we consider the resource conversion from entanglement to quantum coherence in multipartite systems. In Ref. \cite{csrbal16prl}, the authors introduced a task of assisted coherence distillation (ACD) in bipartite systems, where both parties work together to generate the maximal possible coherence on one of the subsystems and the operations are limited to the local quantum incoherent operations and classical communication (LQICC) in which the target party performs local incoherent operations and the assisted party utilizes local quantum operations. It is noted that the task does not limit the output of the coherence in the assisted subsystem and local quantum operations can generate quantum coherence on the subsystem in general (for example, the projection measurement with $\ket{\pm}=(\ket{0}\pm\ket{1})/\sqrt{2}$).

Here, we consider the multi-party local incoherent operations and classical communication (LICC) scenario \cite{str17prx} where all the parties are in the distance labs and each local one can perform only local incoherent operations assisted by classical communication among different parties. In this situation, all of the local parties cannot create quantum coherence and we have the relation $\mbox{LICC}\subset \mbox{LQICC}$ for the two classes of operations due to the local incoherent operation being a subset of local quantum operation.
In the cyclic resource conversion, since the entanglement in multipartite systems comes from quantum coherence via multipartite incoherent operations $\Lambda_I(\rho_A\otimes \sigma_{B_1B_2\cdots B_n})$, the procedure of conversion from entanglement to quantum coherence should be restricted to incoherent operations. Therefore, in the distance lab paradigm, it is desirable to confine the operations to the LICC, which is the free operation for the resource conversion in the whole multipartite system.

In the previous stage from single-party coherence to multipartite entanglement, the generic output state is $\varrho_{AB_1B_2\cdots B_n}=\Lambda_I(\rho_A\otimes \sigma_{B_1B_2\cdots B_n})$ which is not the maximally correlated state (MCS) form under a multipartite incoherent operation assisted by an $N$-partite incoherent ancilla. In this case, the relations on operational entanglement and coherence are characterized by the bounds in Theorem 4. When the local parties want to restore the coherence to the initial party $A$, they can use the protocol of ACD via the LICC \cite{str17prx}, and the corresponding distillable coherence $C_{LICC}^{B_1B_2\cdots B_n|A}$ is upper bounded by its operational coherence $C_d(\varrho_{AB_1B_2\cdots B_n})$ since quantum coherence is monotone under the LICC. Moreover, the ACD via LICC is also upper bounded by the quantum-incoherent (QI) relative entropy \cite{csrbal16prl,str17prx}
\begin{eqnarray}
  C_{LICC}^{B_1B_2\cdots B_n|A}(\varrho_{AB_1B_2\cdots B_n}) &\leq& C_r^{B_1B_2\cdots B_n|A}(\varrho_{AB_1B_2\cdots B_n}) \nonumber\\
   &=& \mbox{min}_{\chi\in \mathcal{QI}} S(\varrho||\chi),
\end{eqnarray}
with $\chi$ being the QI state in the bipartition $A|B_1B_2\cdots B_n$ of multipartite systems.

In the following, we consider the optimal output state under the multipartite incoherent operation $U_{mcn}$ which has the MCS form.
Due to the MCS making the inequalities in Theorem 4 saturated, we obtain the following theorem in the resource conversion from multipartite entanglement to the operational coherence of single-party system.

\textit{Theorem 5}. For the optimal output state under multipartite incoherent operations, its operational entanglements are equal to corresponding multipartite entanglements, which are the upper bounds on the converted coherence via the LICC
\begin{eqnarray}
E_r^M(\rho_{AB_1B_2\dots B_n}^{mc})&=&E_d(\rho_{\alpha|\bar{\alpha}}^{mc})\nonumber\\
&\geq& C_d\left[\Lambda_{LICC}(\rho_{AB_1B_2\dots B_n}^{mc})\right]\nonumber\\
&\geq&  C_d(\rho_r^{LICC}),\\
E_f^{GME}(\rho_{AB_1B_2\dots B_n}^{mc})&=&E_f(\rho_{\alpha|\bar{\alpha}}^{mc})\nonumber\\
&\geq& C_f\left[\Lambda_{LICC}(\rho_{AB_1B_2\dots B_n}^{mc})\right]\nonumber\\
&\geq& C_f(\rho_r^{LICC}),
\end{eqnarray}
where ${\alpha|\bar{\alpha}}$ is an arbitrary bipartite partition in the multipartite systems, $\Lambda_{LICC}$ is the multipartite transformation under the LICC, and $\rho_r^{LICC}=\mbox{Tr}_{\bar{r}}[\Lambda_{LICC}(\rho_{AB_1B_2\dots B_n}^{mc})]$ is the reduced state of the multipartite system with $\bar{r}$ being the traced subsystem.

\textit{Proof}.--- According to Theorem 4, we have the relation $E_d(\rho_{\alpha|\bar{\alpha}}^{mc})=E_r^M(\rho_{AB_1B_2\dots B_n}^{mc})=C_d(\rho_{AB_1B_2\dots B_n}^{mc})$
for the MCS. The first inequality in Eq. (29) is satisfied because the distillable coherence $C_d$ is monotone under the free operation $\Lambda_{LICC}$.
With the relation $C_d(\varrho)=C_r(\varrho)$ and the property that the relative entropy of a state is not increasing after tracing a subsystem out \cite{ved02rmp}
\begin{equation}
S(\mbox{Tr}_p\rho||\mbox{Tr}_p\sigma)\leq S(\rho||\sigma),
\end{equation}
where $\mbox{Tr}_p$ is a partial trace, we can obtain the second inequality in Eq. (29).
Next, we analyze the resource conversion relations in Eq. (30) for which we have
$E_f^{GME}(\rho_{AB_1B_2\dots B_n}^{mc})=E_f(\rho_{\alpha|\bar{\alpha}}^{mc})=C_f(\rho_{AB_1B_2\dots B_n}^{mc})$.
The first inequality is satisfied due to the coherence of formation being monotone under the LICC. Furthermore, via the concavity of von Neumann entropy, we can obtain the entropy of diagonal state $\Delta(\psi_i)$ is not increasing after a partial trace,\textit{ i.e.},
\begin{equation}
S\left(\mbox{Tr}_p\Delta(\psi_i)\right)\leq S\left(\Delta(\psi_i)\right).
\end{equation}
Combining the above property with the convexity of $C_f$, we have the second inequality in Eq. (30), and then the proof  is completed.  \qed

According to Theorem 5, we know that the converted coherence in multipartite systems under a general LICC is upper bounded by the multipartite entanglement in the MCS. Moreover, since the LICC is a subset of the LOCC, the operational quantum entanglement for any initial state in the resource conversion is monotone under the LICC.

When the multipartite system is composed of qubits, the multipartite entanglement indicators $\tau_{MED}$ and $\tau_{MEF}$ are the upper bounds on the converted coherence via the LICC in the arbitrary subsystems,
\begin{eqnarray}
\tau_{MED}(\rho_{AB_1B_2\dots B_n}^{mc})&\geq& C_d(\rho_r^{LICC}),\\
\tau_{MEF}(\rho_{AB_1B_2\dots B_n}^{mc})&\geq& C_f(\rho_r^{LICC}),
\end{eqnarray}
where $\rho_r^{LICC}=\mbox{Tr}_{\bar{r}}[\Lambda_{LICC}(\rho_{AB_1B_2\dots B_n}^{mc})]$ is the reduced state of the multipartite system with $\bar{r}$ being the traced subsystem.

Next, we analyze the resource conversion from multipartite entanglement in the MCS to quantum coherence under the optimal LICC. In this case, the local incoherent operation can be chosen to be the set of Kraus operators $K_j=\ket{j}\bra{\varphi_j}$ with $\ket{\varphi}_j=1/\sqrt{d}\sum_k e^{i\phi_k^j}\ket{k}$ being the mutually orthogonal maximally coherent state and $j=0,1,\dots d-1$ (in fact, all the mutually unbiased bases \cite{mub02alg,mub10ijq,mub13prl} except for the coherence-dependent basis can be used in the incoherent measurement).
After the incoherent measurement $K_j$ is performed on a subsystem, the quantum state of the remained subsystems can be transformed to the MCS form with the help of classical communication (measurement results on the $j$) and an incoherent unitary operation $U_j=\sum_k e^{i\phi_k^j}\proj{k}$ \cite{str17prx}.

Due to the state of remained subsystem being the MCS and having the same nonzero matrix elements, the operational entanglement and coherence are transferred to the subsystem and keep the same amount. Repeating the incoherent measurement and the incoherent unitary operation on all the subsystems $B_i$, the quantum state of remained subsystem A becomes $\rho_A=\sum_{m,n=0}^{d-1} \rho_{mn}\ket{m}\bra{n}$ and the coherence  is restored to the  single-party subsystem. In addition, it is a similar case to convert multipartite entanglement to the coherence of an arbitrary single-party subsystem $B_i$. Thus, we have completed the characterization on resource interconversion between quantum coherence and entanglement in multipartite systems within full incoherent operation scenario.

\subsection{Cyclic resource interconversion between operational coherence and multipartite entanglement within the full incoherent operation scenario}
Recently, a cyclic interconversion between coherence and quantum correlation \cite{modi10prl,ved01jpa,zur01prl} has been investigated experimentally on bipartite optical systems \cite{xiang18prl}, where the assisted coherence distillation via the LQICC is utilized. When the operations are restricted to the LICC, it is desirable to generalize the cyclic conversion to multipartite systems.

\begin{figure}
	\begin{center}
		\epsfig{figure=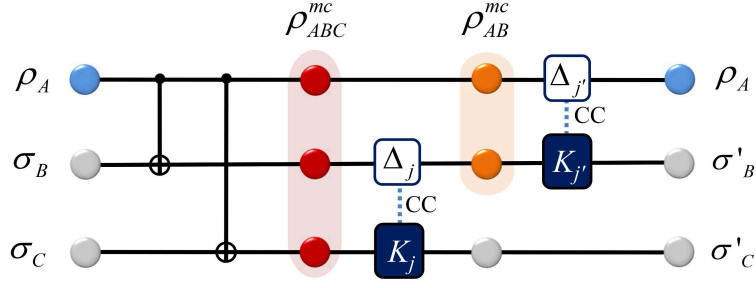,width=0.65\textwidth}
	\end{center}
	\caption{(Color online) A schematic diagram for the cyclic operational resource conversion between single party quantum coherence and genuine tripartite entanglement in three-qubit systems within the full incoherent scenario.}\label{f2}
\end{figure}

In this subsection, we show that the cyclic interconversion can be generalized to operational resource measures for multipartite systems within a full incoherent operation scenario.
As shown in Fig. \ref{f2}, a schematic diagram is given for the cyclic interconversion protocol of  coherence-entanglement-coherence in three-qubit systems without  loss. The initial coherent state is $\rho_A=\sum_{m,n=0}^{1}\rho_{mn}\ket{m}\bra{n}$ and the states of ancillas are $\sigma_B=\sigma_C=\proj{0}$. In the conversion from quantum coherence to entanglement, the optimal multipartite incoherent operation is the generalized tripartite controlled-not gate, which can be realized by two controlled-not gates. Then the output state of the tripartite system $U_{mcn}(\rho_A\otimes\sigma_B\otimes\sigma_C)U_{mcn}^\dagger$ is the maximally correlated state (MCS) and can be expressed as
\begin{equation}\label{35}
\rho_{ABC}^{MI}=\sum_{m,n=0}^{1} \rho_{mn}\ket{mmm}\bra{nnn},
\end{equation}
which is a genuine three-quibt entangled state belonging to the GHZ class under the classification of stochastic local operations and classical comunication (SLOCC) \cite{dur00pra,acin01prl}.
The multipartite entanglements of the MCS are
\begin{eqnarray}\label{36}
E_r^M(\rho_{ABC}^{MI})&=&\tau_{MED}(\rho_{ABC}^{MI})=C_d(\rho_A),\\
E_f^{GME}(\rho_{ABC}^{MI})&=&\tau_{MEF}(\rho_{ABC}^{MI})=C_f(\rho_A).
\end{eqnarray}
Thus we realize the optimal conversion from single-qubit coherence to genuine three-qubit entanglements via multipartite incoherent operation. It is noted that, manipulating the structure of incoherent unitarty operation $U$, one can not change the entangled state from GHZ class to W class. The incoherent unitary operation is restricted to $U=\sum_i \ket{f(i)}\bra{i}$ with $f(i)$ being the one-to-one mapping from the fixed incoherent basis to itself. When the initial  state is a coherent qubit plus two incoherent ancilla, the output state under the multipartite incoherent $U$ has at most two coherent terms in the fixed basis, while the entangled state of W class has three coherent terms $a\ket{100}+b\ket{010}+c\ket{001}$ \cite{acin01prl}.

The tripartite entanglement can be transferred to bipartite systems by an incoherent measurement $\{K_j\}$ with the Kraus operators $K_0=\ket{0}\bra{+}$ and $K_1=\ket{1}\bra{-}$ in which $\ket{\pm}=(\ket{0}\pm\ket{1})/\sqrt{2}$. Via the classical communication ($j=0$ or $1$) between subsystems $B$ and $C$, the observer $B$ performs the corresponding incoherent operation $\Delta_j$ on subsystem $B$ where $\Delta_0=I_2$ and $\Delta_1=\sigma_z$. After the operation $\Delta_j$, the genuine three-qubit entanglement is transferred to two-qubit subsystem and its quantum state is
$\rho_{AB}^{LICC}=\sum_{m,n=0}^{1} \rho_{mn}\ket{mm}\bra{nn}$.
In this case, the operational bipartite entanglement and genuine three-qubit entanglement are connected by equations
$E_r^M(\rho_{ABC})=E_d(\rho_{AB}^{LICC})$ and $E_f^{GME}(\rho_{ABC})=E_f(\rho_{AB}^{LICC})$.
Furthermore, the two-qubit entanglement can be converted to single-party quantum coherence of subsystem $A$ by a set of similar local incoherent operations $\{K_{j\prime}, \Delta_{j\prime}\}$ assisted by the classical communication about the incoherent measurement result $j^\prime=0$ or $1$. After these operations, the quantum state of local subsystem A has the form
\begin{equation}
\rho_A^{LICC}=\sum_{m,n=0}^{1} \rho_{mn}\ket{m}\bra{n},
\end{equation}
which is the same as that of initial single qubit coherent state $\rho_{A}$, and the operational bipartite entanglements are restored to single-qubit operational coherence, \textit{i.e.},
$E_d(\rho_{AB}^{LICC})=C_d(\rho_A^{LICC})$ and $E_f(\rho_{AB}^{LICC})=C_f(\rho_A^{LICC})$. In a similar way, we can choose to convert the entanglement to the single-party coherence in subsystem $B$ when we perform the incoherent measurement on  subsystem $A$.

We have shown that, within a full incoherent operation scenario, the single-qubit coherence and multipartite entanglement can be cyclically interconverted without  loss
\begin{eqnarray}
C_d(\rho_A)&=&E_r^M(\rho_{ABC}^{MI})=C_d(\rho_A^{LICC}),\\
C_f(\rho_A)&=&E_f^{GME}(\rho_{ABC}^{MI})=C_f(\rho_A^{LICC}),
\end{eqnarray}
where the multipartite relative entropy of entanglement and the genuine multipartite entanglement measure based on entanglement of formation are not equal in general since the coherence resource theory is irreversible. The cyclic resource conversion can also be generalized to multipartite multi-level systems.

\section{Genuine multi-level entanglement and resource dynamical properties}

Entanglement of high-dimensional quantum systems is an important resource, which can enhance the capabilities of certain quantum communication protocols \cite{bec00pra,cerf02prl}. There is a kind of high-dimensional bipartite entanglement that cannot be simulated with copies of low-dimensional bipartite systems, which is referred to as genuine multilevel bipartite entanglement \cite{kra18prl}. For example, although two Bell pairs can mimic the test of nonlocality \cite{bru08prl} in a $4\otimes 4$ entangled state, there exist the genuine four-dimensional bipartite entangled states which cannot be decomposed into two Bell-like states \cite{cong17prl,guo18pra}. In the case that the multi-level bipartite entanglement can be simulated with the copies of low-dimensional systems, the entanglement is called as decomposable multi-level entanglement.
Kraft \textit{et al} presented a general theory to characterize the genuine multi-level entanglement \cite{kra18prl}. In particular, for the two-ququart state after the Schmidt decomposition $\ket{\psi}_{AB}=s_0\ket{00}+s_1\ket{11}+s_2\ket{22}+s_3\ket{33}$ (with the assumption $s_0\geq s_1\geq s_2\geq s_3$),
they prove that the entangled state $\ket{\psi}_{AB}$ is decomposable if and only if the determinant of matrix $S=[s_0, s_1;s_2,s_3]$ is zero.

In the resource conversion from quantum coherence to entanglement, the generated quantum state is a multi-level entangled when the input single party state is multi-level coherent. It is an interesting problem that whether or not one can judge the multi-level entanglement property via the initial quantum coherent state. Here, we consider a four-level coherent state
\begin{equation}\label{41}
\ket{\varphi}_{A}=\alpha_0\ket{0}+\alpha_1\ket{1}+\alpha_2\ket{2}+\alpha_3\ket{3},
\end{equation}
where, without loss of generality, we assume that the moduli of amplitudes obey the relation $|\alpha_0|\geq |\alpha_1|\geq |\alpha_2|\geq |\alpha_3|$. In the optimal bipartite resource conversion, the generated bipartite entangled state is a four-level maximally correlated state (MCS) $\rho_{AB}^{mc}=\proj{\varphi}_{AB}$ with $\ket{\varphi}_{AB}=\alpha_0\ket{00}+\alpha_1\ket{11}+\alpha_2\ket{22}+\alpha_3\ket{33}$, and its multi-level entanglement property is related to the coefficients of initial coherent state. After analyzing the determinant of $S$ matrix for the MCS, we can obtain the following observation.

\textit{Observation 1}. The generated two-ququart state $\ket{\varphi}_{AB}$ in the optimal resource conversion is decomposable if and only if the coefficients of initial coherent state satisfy the relation
\begin{equation}\label{42}
|\alpha_0\alpha_3|=|\alpha_1\alpha_2|,
\end{equation}
and the bipartite quantum state is the genuine multi-level entangled when the equality is violated.

We note that, when the initial coherent state can be decomposed into the tensor product state of two qubits $\ket{\varphi}_{A}=(a\ket{0}+b\ket{1})_{A_1}\otimes (c\ket{0}+d\ket{1})_{A_2}$, the equality in Eq. (42) is satisfied and then the generated entanglement in the conversion is decomposable. However, when the initial coherent state is not tensor product, for example, $\ket{\varphi}_{A_1A_2}=\alpha\ket{+}\ket{0}+\beta\ket{-}\ket{1}$, the output entangled state is also decomposable. Therefore, we conclude that the decomposable property of initial coherent state is only a sufficient but not necessary condition for the decomposable property of the generated entangled state in the resource conversion.

Resource dynamics is a fundamental problem in the practical quantum information processing, because quantum systems interact unavoidably with the environment and may lose their quantum coherence or entanglement. In the operational resource conversion between quantum coherence and multipartite entanglement, it is worth analyzing the dynamical behaviors of two kinds of resources under the typical noise environment. Here, we investigate the case of the optimal conversion in three-qubit systems as shown in Fig. \ref{f2}. The initial coherent state is chosen to be $\ket{\psi}_A=\alpha\ket{0}+\beta\ket{1}$ with real coefficients and the output MCS has the form $\ket{\psi}^{mc}_{ABC}=\alpha\ket{000}+\beta\ket{111}$ after the operations of two controlled-not gates. The noise environment we consider is the depolarizing channel $\mathcal{E}(\rho)=pI/d+(1-p)\rho$ \cite{nie00book} with the parameter $p$ being the depolarization probability, under which the coherent state and the converted entangled state are $\rho_A^{\varepsilon}=\mathcal{E}(\ket{\psi}_A\bra{\psi})$ and $\rho_{ABC}^{\varepsilon}=\mathcal{E}(\ket{\psi}_{ABC}^{mc}\bra{\psi})$, respectively. After some calculation, we can obtain the values of two operational coherence measures
\begin{eqnarray}
C_d(\rho_A^{\varepsilon})&=&h(x_1)-h(p/2),\\
C_f(\rho_A^{\varepsilon})&=&h\left[(1+\sqrt{1-x_2^2})/2\right],
\end{eqnarray}
where $h(x)=-x\log_2 x -(1-x)\log_2(1-x)$ is the binary entropy, and the parameters are $x_1=(1-p)\alpha^2+p/2$ and $x_2=2(1-p)\alpha\beta$, respectively.

The distillable entanglement is very difficult to compute in a generic mixed state, but the measure $E_d$ is upper bounded by the logarithmic negativity $E_{\mathcal{N}}(\rho)=\log_2||\rho^{T_A}||_1$ \cite{ple05prl,vid02pra}. On the other hand, Chen \textit{et al} gave a tight lower bound for entanglement of formation in an arbitrary bipartite mixed state, and for the qubit-qudit system, the lower bound can be expressed as a binary entropy function $E_f^{LB}(\rho)=h[(1+\sqrt{1-(\Lambda-1)^2})/2]$ in which $\Lambda$ can be the trace norm $||\rho^{T_A}||_1$ (or the norm of realignment matrix) \cite{chen05prl}.
For the generated entangled state under the depolarizing channel, we can derive the multipartite entanglement indicators via the corresponding bounds of the operational entanglement measures,
\begin{eqnarray}
\tau_{MED}^{UB}(\rho_{ABC}^{\varepsilon})&=&[E_d^{UB}(\rho_{A|BC}^{\varepsilon})^2-\sum_{k\in \{B,C\}} E_d^2(\rho_{Ak}^{\varepsilon})]^{1/2},\nonumber\\
&=&\log_2(1+\zeta/4),\\
\tau_{MEF}^{LB}(\rho_{ABC}^{\varepsilon})&=&[E_f^{LB}(\rho_{A|BC}^{\varepsilon})^2-\sum_{k\in \{B,C\}} E_f^2(\rho_{Ak}^{\varepsilon})]^{1/2}\nonumber\\
&=&h[(\sqrt{\omega}+\sqrt{2-\omega})^2/4],
\end{eqnarray}
where we use the property that the operational entanglement $E_d$ and $E_f$ are zero in two-qubit subsystems, and the parameters are $\zeta=\mbox{max}[0,8(1-p)|\alpha\beta|-p]$, $\omega=||(\rho_{A|BC}^{\varepsilon})^{T_A}||_1$, respectively.

\begin{figure}[t]
	\begin{center}
		\epsfig{figure=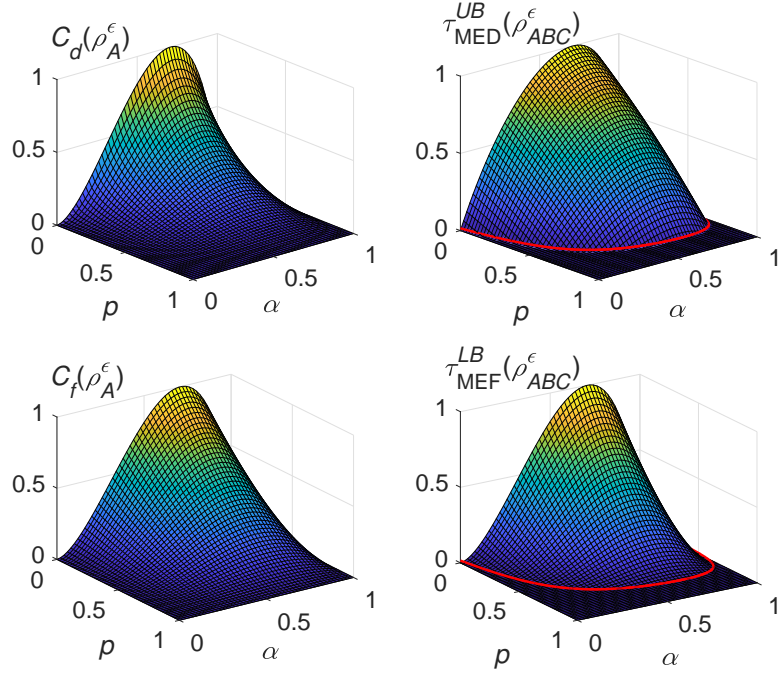,width=0.65\textwidth}
	\end{center}
	\caption{(Color online) The dynamical behaviors of quantum resources along with the depolarizing parameter $p$ and initial state coefficient $\alpha$, where the coherences decay asymptotically but multipartite entanglement experience the ESD (the red solid lines).}\label{f3}
\end{figure}

In Fig. \ref{f3}, we plot the resource measures along with the depolarization parameter $p$ and the state coefficient $\alpha$, where the coherences $C_d$ and $C_f$ decay in an asymptotic way but the multipartite entanglements $\tau_{MED}^{UB}$ and $\tau_{MEF}^{LB}$ disappear in finite time (\textit{i.e.}, entanglement sudden death, ESD). After some calculation, we have the ESD lines for the multipartite entanglement in the figure, which have the same expression,
\begin{equation}
p=\frac{8\alpha\sqrt{1-\alpha^2}}{1+8\alpha\sqrt{1-\alpha^2}}.
\end{equation}
These behaviors coincide with the previous study on quantum correlation dynamics \cite{zyc01pra,sch03jmo,yu06prl,yu09sci}, where quantum systems experience disentanglement in a finite time under the noise environment, even if their coherence is lost asymptotically.

\section{Discussion and conclusion}

In the resource conversion from coherence to entanglement, we restrict the operation to multipartite incoherent operation \cite{yao15pra} which is different from the bipartite case \cite{str15prl,xiang18prl}. For the generated multipartite entangled state, its distillable entanglement and entanglement of formation in an arbitrary bipartite partition $\alpha|\bar{\alpha}$ are upper bounded by the corresponding operational single-party coherences. Furthermore, the output state is multipartite entangled in general, and we have established a set of  relationships between single-party coherence and multipartite entanglement. In particular, we have shown that the multipartite generalized controlled-not operation $U_{mcn}$ is the optimal operation which makes all the resource conversion relationships saturated.

The operational framework between single-party coherence and multipartite entanglement can be further generalized. When the initial state is chosen to be a multipartite coherent state $\rho_{AB_1B_2\cdots B_n}$, the similar relations like those in Theorems 1-3 and Corollary 1 are still satisfied where the distillable coherence and coherence of formation are replaced with $C_d(\rho_{AB_1B_2\cdots B_n})$ and $C_f(\rho_{AB_1B_2\cdots B_n})$, and the bipartite or multipartite entanglement $E[\Lambda_I(\rho_{A}\otimes \sigma_{B_1B_2\dots B_n})]$ is changed to $E[\Lambda_I(\rho_{AB_1B_2\dots B_n})]$. That is to say, the coherence of the whole initial multipartite systems will set the upper bound over the converted entanglements. In a specific case that the $B$ subsystem starts off with some amount of coherence and the initial state is chosen to be $\rho_{A}\otimes \rho_{B_1B_2\cdots B_n}$, the converted bipartite and multipartite entanglements via multipartite incoherent operations
are bounded by $C(\rho_A)+C(\rho_{B_1B_2\cdots B_n})$ due to the additive property of $C_r$ and $C_f$ \cite{win16prl}. However, in the above cases, the multipartite generalized controlled-not operation $U_{mcn}$ is not the optimal multipartite incoherent operation, and the exploration for the optimal operation is still an open problem need to be studied in future.

In a generic $N$-partite quantum state $\rho_{A_1A_2\cdots A_n}$, we prove that the multipartite entanglements $E_r^M$ and $E_f^{GME}$ are not larger than its operational coherences $C_d$ and $C_f$, and the bounds are saturated for the MCS.
Moreover, the conversion from multipartite entanglement of the MCS to single-party coherence is restricted to the LICC \cite{str17prx}, which makes the cyclic resource conversion within a full incoherent operation scenario. In contrast to the LQICC protocol \cite{chi16prl,xiang18prl}, the LICC scheme keeps the coherence of whole multipartite system from increasing. In Sec. IIC, we presented the scheme of cyclic resource interconversion in three-qubit systems where the single party coherence, bipartite entanglement and genuine tripartite entanglement can be freely interconverted without  loss.

The cyclic resource conversion without loss \cite{xiang18prl} can provide potential flexibilities on utilizing quantum coherence or multipartite entanglement to perform certain tasks in quantum information processing, where one may obtain the operational benefits from the resource conversion. Moreover, the cyclic conversion makes quantum coherence and multipartite entanglement be compared quantitatively under the full incoherent operation scenario, which is useful to give a unified characterization on the two kinds of resources in a certain operational framework.

High-dimensional entanglement can enhance the capabilities of quantum communication protocols \cite{bec00pra,cerf02prl}. In the optimal resource conversion from coherence to entanglement, we have provided a method for detecting genuine bipartite multi-level entanglement \cite{kra18prl} via the coefficients of initial coherent state. This method can be further generalized to multipartite case. For an $N$-partite output state $\ket{\psi_N}$ in the optimal conversion, we can judge whether it is  genuine bipartite multi-level entangled by the initial coherent coefficients. When $\ket{\psi_N}$ is not decomposable in an arbitrary partition $\alpha|\bar{\alpha}$, we can obtain that the quantum state is genuine multipartite multi-level entangled \cite{kra18prl}. It should be noted that the decomposability of initial coherent state is not equivalent to that of the generated multi-level entangled state.

In conclusion, we have established a set of resource conversion relationships between coherence and entanglements in multipartite systems within a full incoherent operation scenario, where the operational resource measures and related multipartite quantifiers are focused. Via the multipartite incoherent operation and the assisted coherence distillation by LICC where coherence of the global state is not a freely available resource, we can realize the interconversion between single-party coherence and multipartite entanglement. Moreover, through the procedure of resource conversion, we have been able to bridge the coherent states and genuine multi-level entangled states by the initial coherent coefficients, and to analyze the asymptotical decay of coherence and ESD behavior of multipartite entanglement in a noise environment. By uncovering the operational connection between coherence and  entanglement in multipartite systems, the present work provides a set of useful tools for quantum resource theory in many-body systems.

\section*{Acknowledgments}
Y.-K.B. would like to thank M.-Y. Ye and M.-L. Hu for many useful discussions.
This work was supported by the URC fund of HKU, NSF-China (Grants Nos. 11575051, 11947090, and 11904078), Hebei NSF (Grants Nos. A2016205215, A2019205263 and A2019205266), and project of China Postdoctoral Science Foundation (Grant No. 2020M670683). J. R. and L.-H. R. were also supported by the funds of Hebei Education Department (Grant No. QN2019092) and Hebei Normal University (Grants Nos. L2018B02 and L2019B07).

\appendix
\section{The optimal multipartite incoherent operation}

The coherence resource theory formulated by Baumgratz \textit{et al} \cite{bcp14prl} can be extended into multipartite scenario \cite{yao15pra}. The $N$-partite incoherent state has the form
\begin{equation}
\sigma_n=\sum_{\vec{i}} p_{\vec{i}} \proj{\vec{i}},
\end{equation}
where $p_{\vec{i}}$ are probabilities and $\ket{\vec{i}}=\ket{i_1}\otimes \ket{i_2}\otimes \cdots \ket{i_n}$ with $\ket{i_k}$ being a pre-fixed local basis of the $k$th subsystem. The $N$-partite incoherent operation can be described by a completely positive trace preserving map $\Lambda_I$ which has the form
\begin{equation}\label{A2}
\Lambda_I(\rho)=\sum_l K_l\rho K_l^\dagger,
\end{equation}
where the set of Kraus operators $\{K_l\}$ satisfies the properties $\sum_l K^\dagger_lK_l=I$ and $K_l\mathcal{I}K_l^\dagger\subset \mathcal{I}$ with $\mathcal{I}$ being now the set of $N$-partite incoherent states.

Below Eq. (24) of the main text, we pointed out that the multipartite generalized controlled-not operation $U_{mcn}$ is the optimal multipartite incoherent operation in the conversion from quantum coherence to entanglement  in multipartite systems. Here, we first show that $U_{mcn}$ is a multipartite incoherent operation. In this case, we consider that the incoherent state $\sigma_{AB_1\dots B_n}$ is  $(N+1)$-partite, where the pre-fixed basis is $\{\ket{i}\otimes \ket{j_1}\otimes \dots\otimes\ket{j_n}\}$ with $\ket{i}$ being local basis of subsystem $A$ and $\ket{j_k}$ local basis of the $k$th subsystem $B_k$.
The multipartite generalized controlled-not operation is unitary and can be written as
\begin{eqnarray}
U_{mcn}=&&\sum_{i=0}^{d-1}\sum_{j_1=0}^{d-1}\ldots\sum_{j_n=0}^{d-1}\ket{i}\bra{i}\otimes\ket{(i+j_1)\mbox{mod}~d}\bra{ j_1}\otimes\ldots\nonumber\\
&&\otimes\ket{(i+j_n)\mbox{mod}~d}\bra{j_n},
\end{eqnarray}
where the dimensions of all the local systems are equal and the operation can realize the transformation $\ket{i}\ket{j_1}\dots\ket{j_n}\rightarrow \ket{i}\ket{(i+j_1)\mbox{mod}~d}\dots \ket{(i+j_n)\mbox{mod}~d}$. Moreover, when the dimension $d_k$ of subsystem $B_k$ is larger than that of subsystem $A$, the transformation $U_{mcn}$ does not change the prefixed basis for $j_k\geq d$. In the case that the dimension $d_k$ of subsystem $B_k$ is less than that of subsystem $A$, one can add an extra ancilla $B_k^\prime$ to the subsystem $B_k$, which can enlarge the Hilbert space of the new subsystem $\tilde{B_k}=B_k B_k^\prime$.
Therefore, the operation $U_{mcn}$ maps the set of pre-fixed basis of multipartite systems into itself, and then it is a multipartite incoherent operation  satisfying the property shown in Eq. (\ref{A2}).

Next, we analyze the optimality of the multipartite generalized controlled-not operation $U_{mcn}$ in the operational resource conversion of multipartite systems. In Theorem 1 of the main text, we prove that the operational coherences of single party $A$ are not less than the generated entanglement of multipartite systems $AB_1\cdots B_n$ under a multipartite incoherent operation $\Lambda_I$. When the initial state $\rho_{AB_1\cdots B_n}$ is a single-party coherent state $\rho_A=\sum_{mn}\rho_{mn}\ket{m}\bra{n}$ accompanied by the ancillary state $\sigma_{B_1B_2\cdots B_n}=\proj{0}^{\otimes n}$, we can choose the multipartite incoherent operation $\Lambda_I$ to be $U_{mcn}$. Then the output state is
\begin{eqnarray}
\Lambda_I\left(\rho_{AB_1\cdots B_n}\right)&=&U_{mcn}\left(\rho_A\otimes\proj{0}^{\otimes n}_{B_1\cdots B_n}\right)U_{mcn}^{\dagger}\nonumber\\
&=&\sum_{m,n=0}^{d-1} \rho_{mn}\ket{mm\ldots m}\bra{nn\ldots n}\nonumber\\
&=&\rho_{AB_1\cdots B_n}^{mc}
\end{eqnarray}
where $\rho_{AB_1\cdots B_n}^{mc}$ is the MCS in the $(N+1)$-partite systems. Since the MCS has the properties
$E_d(\rho_{\alpha|\bar{\alpha}}^{mc})=C_d(\rho_{A})$ and $E_c(\rho_{\alpha|\bar{\alpha}}^{mc})=E_f(\rho_{\alpha|\bar{\alpha}}^{mc})=C_f(\rho_{A})$ with $\alpha|\bar{\alpha}$ being an arbitrary bipartite partition in the multipartite systems, we can obtain that the inequalities in Theorem 1 and Corollary 1 are saturated and the operation $U_{mcn}$ is the optimal multipartite incoherent operation in the resource conversion.

It should be noted that the multipartite incoherent operation is different from that of the bipartite case. Therefore, the results in Theorem 1 is not a trivial extension in which the subsystems $B_1B_2\cdots B_n$ are regarded as a whole system $B$. For example, when the input state is a single-qubit maximally coherent state $\ket{+}_A=(\ket{0}+\ket{1})/\sqrt{2}$ accompanied by two ancillary qubits $\ket{00}_{B_1B_2}$, the output state under the optimal tripartite incoherent operation (tripartite generalized controlled-not gate) is a GHZ state $(\ket{000}_{AB_1B_2}+\ket{111}_{AB_1B_2})/\sqrt{2}$, which is a genuine tripartite entangled state and makes the inequalities in Theorem 1 saturated in an arbitrary partition such as $AB_1|B_2$ and $AB_2|B_1$.
Meantime, the multipartite entanglement inequalities in Corollary 1 are saturated too.
However, when the input state is $\ket{+}_A\ket{0}_B$ in $2\otimes 4$ systems, the output state for the optimal bipartite incoherent operation (bipartite controlled-not gate) is a bipartite Bell state $(\ket{00}_{AB}+\ket{11}_{AB})/\sqrt{2}$.

\section{Monogamy property of distillable entanglement in multi-qubit systems}
In Eq. (18) of the main text, we define the multipartite entanglement indicator
\begin{equation}
 \tau_{MED}(\rho_{N})=[\max\{0,E_d^2(\rho_{A_1|\bar{A_1}})-\sum_{i=2}^n E_d^2(\rho_{A_1A_i})\}]^{\frac{1}{2}},
\end{equation}
which is based on the distribution of squared distillable entanglement. Entanglement monogamy is an important property in many-body quantum systems and the residual entanglement can characterize the genuine multipartite entanglement in the composite system.

We first analyze the monogamy relation of the squared distillable entanglement in an $N$-qubit pure state $\ket{\psi}_N=\ket{\psi}_{A_1A_2\cdots A_n}$.  In this case, the multipartite entanglement indicator $\tau_{MED}(\rho_{N})$ is effective  and we  have the property
\begin{eqnarray}
&&E_d^2(\ket{\psi}_{A_1|A_2\cdots A_n})-\sum_{i=2}^n E_d^2(\rho_{A_1A_i})\nonumber\\
&&\geq E_f^2(\ket{\psi}_{A_1|A_2\cdots A_n})-\sum_{i=2}^n E_f^2(\rho_{A_1A_i})\nonumber\\
&&\geq 0,
\end{eqnarray}
where $E_d(\ket{\psi}_{A_1|A_2\cdots A_n})$ quantifies bipartite distillable entanglement in the $N$-qubit system and $E_d(\rho_{A_1A_i})$ two-qubit distillable entanglement,
the relations $E_d(\ket{\psi}_{A_1|A_2\cdots A_n})=E_f(\ket{\psi}_{A_1|A_2\cdots A_n})$ and $E_f(\rho_{A_1A_i})\geq E_d(\rho_{A_1A_i})$ are used in the first inequality, and the monogamy property of $E_f^2$ \cite{bxw14prl} is utilized in the second inequality.

For the case of mixed states $\rho_{A_1A_2\cdots A_n}$, it is still an open problem that whether the distillable entanglement is monogamous. However, when the residual entanglement is positive, namely, $E_d^2(\ket{\psi}_{A_1|A_2\cdots A_n})-\sum_{i=2}^n E_d^2(\rho_{A_1A_i})> 0$ , the indicator $\tau_{MED}$ can indicate that there exists genuine multipartite entanglement which cannot be stored in two-qubit subsystems. For example, in the optimal resource conversion from coherence to entanglement, the output state has the form $\rho_{N+1}^{mc}=\rho_{AB_1B_2\cdots B_n}^{mc}=\sum_{m,n=0}^{d-1} \rho_{mn}\ket{mm\ldots m}\bra{nn\ldots n}$. Since all the two-qubit reduced states of $\rho^{mc}_{N+1}$ are separable, we have
\begin{eqnarray}
\tau_{MED}(\rho^{mc}_{N+1})&=&[E_d^2(\rho^{mc}_{A|B_1B_2\cdots B_n})-\sum_{i=1}^n E_d^2(\rho_{A_1B_i})]^{1/2}\nonumber\\
&=&E_d(\rho^{mc}_{A|B_1B_2\cdots B_n}),
\end{eqnarray}
which indicates the existence of the genuine multipartite entanglement in multi-qubit systems.
Moreover, it is a similar situation for the entanglement dynamics under depolarizing noise that the nonzero $\tau_{MED}^{UB}(\rho_{ABC}^{\varepsilon})$ indicates the genuine three-qubit entanglement as shown in Fig. \ref{f3} of the main text.

\section{The optimality of $U_{mcn}$ for multipartite entanglement conversion}

In Theorem 2 of the main text, when the initial single-party coherent state $\rho_A=\sum \rho_{mn}\ket{m}\bra{n}$ is accompanied by
ancilla state being in $\ket{00\cdots 0}_{B_1B_2\cdots B_n}$, the output state under multipartite generalized controlled-not operation $U_{mcn}$ is
\begin{equation}\label{C1}
\rho_{AB_1B_2\dots B_n}^{mc}=\sum_{m,n=0}^{d-1} \rho_{mn}\ket{mm\ldots m}\bra{nn\ldots n}.
\end{equation}
For this output state, the multipartite relative entropy of entanglement has the property
\begin{eqnarray}\label{C2}
E_r^M(\rho_{AB_1B_2\dots B_n}^{mc})&=&S(\rho_{AB_1B_2\dots B_n}^{mc}||\sigma_s^{N+1})\nonumber\\
&=& S(\rho_{AB_1B_2\dots B_n}^{mc}||\sigma_s^{A|B_1B_2\cdots B_n})\nonumber\\
&\geq& E_r(\rho^{mc}_{A|B_1B_2\cdots B_n})\nonumber\\
&\geq& S(\rho_A^d)-S(\rho^{mc}_{AB_1B_2\cdots B_n})\nonumber\\
&=& C_d(\rho_A),
\end{eqnarray}
where in the first equality we use the nearest $(N+1)$-partite separable state $\sigma_s^{N+1}$ for multipartite relative entropy of entanglement, in the second equality we cut the $(N+1)$-partite separable state into bipartite partition $A|B_1B_2\cdots B_n$ which results in the relative entropy being not less than the bipartite relative entropy of entanglement, in the second inequality we use the lower bound for $E_r(\rho)$ \cite{hhh00prl,dw05prsa}, and in the last equality we use the relation $S(\rho^{mc}_{AB_1B_2\cdots B_n})=S(\rho_A)$ and the definition of the distillable coherence.
Combining the relation $C_d(\rho_A)\geq E_r^M[\Lambda_I(\rho_A\otimes\sigma_{B_1B_2\cdots B_n})]$ in Theorem 2 with Eq. (\ref{C2}), we have
\begin{equation}
E_r^M(\rho_{AB_1B_2\dots B_n}^{mc})=C_d(\rho_A),
\end{equation}
and then the multipartite generalized controlled-not operation $U_{mcn}$ is the optimal multipartite incoherent operation. Next, we consider the GME based on entanglement of formation, for which the value of $\rho_{AB_1B_2\dots B_n}^{mc}$ under optimal pure state decomposition $\{p_i,\ket{\psi_i}\}$ is
\begin{eqnarray}
E_f^{GME}(\rho_{AB_1B_2\cdots B_n}^{mc})&=&\sum_{i}p_i E_f^{GME}(\ket{\psi_i})\nonumber\\
&=&\sum_{i}p_i E_f(\ket{\psi_i}_{\alpha|\bar{\alpha}})\nonumber\\
&=&\sum_{i}p_i E_f(\ket{\psi_i}_{A|B_1B_2\cdots B_n})\nonumber\\
&=&\sum_{i}p_i C_f(\ket{\psi_i}_{AB_1B_2\cdots B_n})\nonumber\\
&=&C_f(\rho_{AB_1B_2\cdots B_n}^{mc})\nonumber\\
&=&C_f(\rho_A),
\end{eqnarray}
where in the second equality we use the minimal entanglement in bipartite partition $\alpha|\bar{\alpha}$,
in the third equality we use the property that any $\ket{\psi_i}$ in the support of the MCS has the form $\ket{\psi_i}=\sum_j q_j\ket{jj\cdots j}$ with $\sum_j|q_j|^2=1$ and the minimal entanglement can choose the partition $A|B_1B_2\cdots B_n$, in the fourth equality the property $E_f(\ket{\psi_i})=C_f(\ket{\psi_i})$, and the last equality holds since $\rho_{AB_1B_2\cdots B_n^mc}$ and $\rho_A$ have the same nonzero matrix elements. Therefore, we obtain that the multipartite generalized controlled-not operation $U_{mcn}$ is optimal in the resource conversion from single-party coherence of formation to the GME based on entanglement of formation.

In Theorem 3 of the main text, we prove that, in the resource conversion via multipartite incoherent operation, the operational coherences of single-party system are not less than the multipartite entanglement indicators based on operational entanglements. When we choose the operation $U_{mcn}$, the output state is the MCS state in Eq. (\ref{C1}) for which its multipartite entanglement indicators are
\begin{eqnarray}
\tau_{MED}&=&E_d(\rho_{A|B_1B_2\cdots B_n}^{mc})=C_d(\rho_A),\\
\tau_{MEF}&=&E_f(\rho_{A|B_1B_2\cdots B_n}^{mc})=C_f(\rho_A),
\end{eqnarray}
where we use the property of MCS that its two-qubit reduced state $\rho_{AB_i}$ is separable, and the operational entanglements of MCS are equal to the operational coherences. Such that the $U_{mcn}$ is optimal.


\begin{thebibliography}{99}
\bibitem{loccnote}  Two-way LOCC operations can be written as $\Gamma_1(\varrho_{AB})=\sum_i(A_i\otimes I\varrho_{AB}A_i^{\dagger}\otimes I)\otimes \proj{i}_{B^{\prime}}$ and $\Gamma_2(\varrho_{AB})=\sum_i(I\otimes B_i\varrho_{AB}I\otimes B_i^{\dagger})\otimes \proj{i}_{A^{\prime}}$ in which $\sum_iA_i^{\dagger}A_i=I_A$ and $\sum_iB_i^{\dagger}B_i=I_B$, and the form of other class of LOCC can see Sec. XIB in Ref. [3].
\end{thebibliography}
\end{document}